\pdfoutput=1

\documentclass[11pt,twoside,a4paper,cmspaper,final,collab]{cms-tdr}
\def\svnVersion{435966P}\def\cmsCernNoTag{CERN-EP-2017-113}\def\cmsCernDate{\today}\def\cmsMessage{Published in the Journal of High Energy Physics as \href{http://dx.doi.org/10.1007/JHEP11(2017)029}{\doi{10.1007/JHEP11(2017)029.}}}

\begin{document}\cmsNoteHeader{SUS-16-043}

\hyphenation{had-ron-i-za-tion}
\hyphenation{cal-or-i-me-ter}
\hyphenation{de-vices}
\RCS$Revision: 426565 $
\RCS$HeadURL: svn+ssh://svn.cern.ch/reps/tdr2/papers/SUS-16-043/trunk/SUS-16-043.tex $
\RCS$Id: SUS-16-043.tex 426565 2017-09-22 09:04:06Z mliu $
\newlength\cmsFigWidth
\ifthenelse{\boolean{cms@external}}{\setlength\cmsFigWidth{0.85\columnwidth}}{\setlength\cmsFigWidth{0.4\textwidth}}
\ifthenelse{\boolean{cms@external}}{\providecommand{\cmsLeft}{top\xspace}}{\providecommand{\cmsLeft}{left\xspace}}
\ifthenelse{\boolean{cms@external}}{\providecommand{\cmsRight}{bottom\xspace}}{\providecommand{\cmsRight}{right\xspace}}
\providecommand{\NA}{\ensuremath{\text{---}}\xspace}

\newcommand{\fullLumi}{35.9\fbinv}
\newcommand{\mt}{\ensuremath{M_{\mathrm{T}}}\xspace}
\newcommand{\mbb}{\ensuremath{M_{\bbbar}}\xspace}
\newcommand{\mjj}{\ensuremath{M_{\mathrm{jj}}}\xspace}
\newcommand{\mct}{\ensuremath{M_{\mathrm{CT}}}\xspace}
\newcommand{\ttsl}{\ensuremath{\ttbar\to\ell+\text{jets}}\xspace}
\newcommand{\wjets}{\ensuremath{\PW+\text{jets}}\xspace}
\newcommand{\wl}{\ensuremath{\PW+\mathrm{LF}}\xspace}
\newcommand{\whf}{\ensuremath{\PW+\mathrm{HF}}\xspace}
\newcommand{\wzbb}{\ensuremath{\PW\Z\to\ell\nu\bbbar}\xspace}
\newcommand{\zjets}{\ensuremath{\Z+\text{jets}}\xspace}
\newcommand{\gjets}{\ensuremath{\gamma+\text{jets}}\xspace}
\newcommand{\ttw}{\ensuremath{\ttbar\PW}\xspace}
\newcommand{\ttz}{\ensuremath{\ttbar\Z}\xspace}
\newcommand{\whbb}{\ensuremath{\PW\PH \to \ell\nu \bbbar}\xspace}
\newcommand{\chipmo}{\PSGcpmDo\xspace}
\newcommand{\chitn}{\PSGczDt\xspace}
\newcommand{\lsp}{\PSGczDo\xspace}
\newcommand{\crdl}{\ensuremath{\mathrm{CR}2\ell}\xspace}
\newcommand{\crzb}{\ensuremath{\mathrm{CR}0\PQb}\xspace}
\newcommand{\crmbb}{\ensuremath{\mathrm{CRM}\bbbar}\xspace}

\cmsNoteHeader{SUS-16-043}
\title{Search for electroweak production of charginos and neutralinos in WH events in proton-proton collisions at $\sqrt{s}=13\TeV$}

\date{\today}

\abstract{
Results are reported from a search for physics beyond the standard model in proton-proton collision events with a charged lepton (electron or muon), two jets identified as originating from a bottom quark decay, and significant imbalance in the transverse momentum. The search was performed using a data sample corresponding to 35.9\fbinv, collected by the CMS experiment in 2016 at $\sqrt{s} = 13\TeV$. Events with this signature can arise, for example, from the electroweak production of gauginos, which are predicted in models based on supersymmetry. The event yields observed in data are consistent with the estimated standard model backgrounds. Limits are obtained on the cross sections for chargino-neutralino ($\chipmo\chitn$) production in a simplified model of supersymmetry with the decays $\chipmo \to\PW^{\pm} \lsp$ and $\chitn\to\PH \lsp$. Values of $m_{\chipmo}$ between 220 and 490\GeV are excluded at 95\% confidence level by this search when the $\lsp$ is massless, and values of $m_{\lsp}$ are excluded up to 110\GeV for $m_{\chipmo} \approx  450$\GeV.
}

\hypersetup{%
pdfauthor={CMS Collaboration},%
pdftitle={Search for electroweak production of charginos and neutralinos in WH events in proton-proton collisions at sqrt(s) =13 TeV},%
pdfsubject={CMS},%
pdfkeywords={CMS, physics, SUSY}}

\maketitle

\section{Introduction}

Supersymmetry (SUSY)~\cite{Ramond:1971gb,Golfand:1971iw,Neveu:1971rx,Volkov:1972jx,Wess:1973kz,Wess:1974tw,Fayet:1974pd,Nilles:1983ge}
is a theoretically attractive extension of the standard model (SM) that is based on a symmetry between bosons and fermions.
SUSY predicts the existence of a superpartner for every SM particle, with the same gauge quantum numbers but differing by one half unit of spin.
In R-parity conserving SUSY models, supersymmetric particles are created in pairs, and the lightest supersymmetric particle (LSP) is
stable~\cite{FARRAR1978575,djoua,carena}. As a result, SUSY also provides a potential connection to cosmology as the LSP, if neutral and stable, may be a
viable dark matter candidate.

Previous searches based on 13\TeV proton-proton collision data at the CERN LHC focused on strong production of colored SUSY
particles~\cite{ATLASat13TeV:manyjets,ATLASat13TeV:monojet,ATLASat13TeV:hadronic,ATLASat13TeV:manyb,MT2at13TeV,Sirunyan:2017cwe,RAZORat13TeV,AlphaTat13TeV,Aad:2016tuk,Sirunyan:2017uyt,Aaboud:2016lwz,Aad:2016jxj,Aad:2016qqk,Khachatryan:2017rhw,Khachatryan:2016xdt,Khachatryan:2017qgo,Khachatryan:2016uwr}.
Pair production of these particles would have the largest cross section for SUSY processes and therefore provides the strongest discovery potential with small datasets.
However, the absence of signals in these searches suggests that strongly produced SUSY particles may be too massive to be found with the present data.
In contrast, neutralinos (\PSGcz) and charginos (\PSGcpm), mixtures of the superpartners of the SM electroweak gauge
bosons and the Higgs bosons, can have masses within the accessible range. Because of the absence of color charge,
the production cross sections are lower, and these particles may have thus far eluded detection.
This provides strong motivation for dedicated searches for electroweak SUSY particle production.

Depending on the mass spectrum, the charginos and neutralinos can have significant decay branching fractions to
vector bosons V ($\PW$ or \Z) and the Higgs boson (\PH).
Here, ``\PH'' refers to the 125\GeV Higgs boson~\cite{Khachatryan:2016vau}, interpreted as the lightest CP-even state of an extended Higgs sector.
The \PH boson is expected to have SM-like properties if all of the other Higgs bosons are much heavier~\cite{Martin:1997ns}.
The observation of a Higgs boson in a SUSY-like process would provide evidence that SUSY particles couple to the Higgs field,
a necessary condition for SUSY to stabilize the Higgs boson mass.
Pair production of neutralinos and/or charginos can thus lead to the $\PH\PH$, V\PH, and VV decay modes,
with a large fraction of the possible final states containing at least one isolated lepton.
Such events can be easily selected with simple triggers and do not suffer from large quantum chromodynamics multijet background.

\begin{figure}[hbt]
  \centering
        \includegraphics[width=0.4\linewidth]{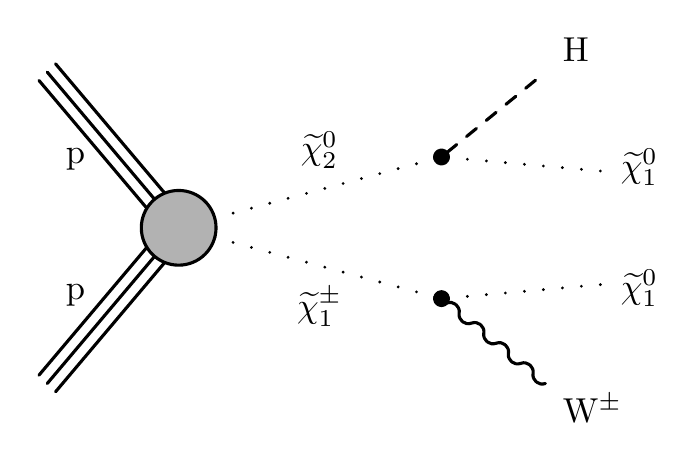}
    \caption{
      Diagram corresponding to the SUSY simplified model targeted by this analysis, \ie,
      chargino-neutralino production, with the chargino decaying to a $\PW$ boson and an LSP,
      while the heavier neutralino decays to a Higgs boson and an LSP.
\label{fig:tchiwh}
}

\end{figure}

In this paper we focus on a simplified model~\cite{bib-sms-1,bib-sms-2,bib-sms-3,bib-sms-4,Chatrchyan:2013sza}
of supersymmetric chargino-neutralino ($\chipmo\chitn$) production with the decays
$\chipmo\to\PW^{\pm}\lsp$ and $\chitn\to\PH\lsp$, as shown in Fig.~\ref{fig:tchiwh}.
Both the $\chipmo$ and $\chitn$ are assumed to be wino-like and have the same mass.
The lightest neutralino $\lsp$, produced in the decays of $\chipmo$ or the $\chitn$, is considered to be the stable LSP, which escapes
detection.
When the $\PW$ boson decays leptonically,
this process typically results in a signature with one lepton, two jets that originate from the decay $\PH\to\bbbar$, and
large missing transverse momentum from the neutrino in the $\PW$ boson decay and the LSPs.

Results of searches for electroweak pair production of SUSY particles were previously reported by the ATLAS and CMS
Collaborations using data sets of 8\TeV proton-proton (\Pp\Pp) collisions \cite{Khachatryan:2014mma,Khachatryan:2014qwa,Aad:2015eda} in a variety of
event topologies and final states. No excesses above the SM expectations were observed, and the results
of those searches were used to place lower limits on the mass of pair-produced charginos and neutralinos. Assuming
mass-degenerate $\chipmo$ and $\chitn$, and sleptons (the SUSY partners of the SM leptons) with lower masses, the
searches probed masses up to approximately 700\GeV.
For the $\PW\PH$ decays assumed here, the strongest mass limit was around 270\GeV.
With the increase of the LHC collision energy from 8 to 13\TeV,
and a significantly larger data set, searches based on 13\TeV data have the potential to quickly surpass the sensitivity of the
previous analyses.

This paper presents the result of a search using a data set corresponding to an integrated luminosity of \fullLumi of $\Pp\Pp$ collisions
collected at a center-of-mass energy of 13\TeV with the CMS detector in 2016.
The results are interpreted in the simplified SUSY model with chargino-neutralino production depicted in Fig.~\ref{fig:tchiwh}.

\section{The CMS detector}
\label{sec:detector}

The central feature of the CMS apparatus is a superconducting solenoid, 13\unit{m} in length and 6\unit{m} in diameter, which
provides an axial magnetic field of 3.8~T. Within the field volume are several particle detection systems.
Charged-particle trajectories are measured with silicon pixel and strip trackers, covering $0 \leq \phi < 2\pi$ in
azimuth and $\abs{\eta} < 2.5$ in pseudorapidity, where $\eta \equiv -\ln [\tan (\theta/2)]$ and $\theta$ is the
polar angle of the trajectory of the particle with respect to the counterclockwise beam direction.
The transverse momentum, the component of the momentum $p$ in the plane orthogonal to the beam, is defined in
terms of the polar angle as $\pt = p \sin \theta$.
A lead-tungstate crystal electromagnetic calorimeter and a brass and scintillator hadron calorimeter surround
the tracking volume, providing energy  measurements of electrons, photons, and hadronic jets
in the range $\abs{\eta} < 3.0$.
Muons are identified and measured within $\abs{\eta} < 2.4$ by gas-ionization detectors embedded in the steel
flux-return yoke of the solenoid.
Forward calorimeters on each side of the interaction point encompass~$3.0 < \abs{\eta} < 5.0$.
The detector is nearly hermetic, allowing momentum imbalance measurements in the plane transverse to the beam direction.
A two-tier trigger system selects pp collision events of interest for use in physics analyses.
A detailed description of the CMS detector, together with a definition of the coordinate system
used and the relevant kinematic variables, can be found in Ref.~\cite{JINST}.

\section{Event samples, reconstruction, and selection}

\subsection{Object definition and preselection}
\label{sec:objdef}

Event reconstruction is based on the particle-flow (PF) algorithm~\cite{CMS-PAS-PFT-09-001,CMS-PAS-PFT-10-001},
which combines information from the tracker, calorimeter, and muon systems to reconstruct and identify PF candidates,
\ie, charged and neutral hadrons, photons, muons, and electrons.
To select collision events, we require at least one
reconstructed vertex.
The reconstructed vertex with the largest value of summed physics-object
$\pt^2$ is taken to be the primary $\Pp\Pp$ interaction vertex. The physics
objects are the objects returned by a jet finding
algorithm~\cite{antikt,FastJet} applied to all charged tracks
associated with the vertex, plus the corresponding associated missing transverse momentum.
The missing transverse momentum vector, \ptvecmiss, is defined as the negative vector sum of the momenta of all
reconstructed PF candidates projected onto the plane perpendicular to the proton beams. Its magnitude is referred to as \ETm.
Events with possible contributions from beam halo processes
or anomalous noise in the calorimeter can have large values of \ETm and are rejected using dedicated filters~\cite{Chatrchyan:2011tn}.

Data events are selected using triggers that require the presence of an isolated electron or muon with \pt
thresholds of 27\GeV or 24\GeV, respectively.
Muon events may also be accepted using a trigger that does not require isolation but instead requires $\pt >$~50\GeV.
The trigger efficiency, measured using a data sample of $\ensuremath{\cPZ/\Pgg^\star} \to \ell \ell$
events, varies in the range 70--95\% (85--92\%)
depending on the $\eta$ and \pt of the electron (muon).

Selected events are required to have exactly one lepton (electron or muon), with
electrons (muons) satisfying
$\pt > 30 (25)$\GeV and $\abs{\eta} < 1.44 (2.1)$.
Electron candidates are reconstructed starting from a cluster of energy deposits in the
electromagnetic calorimeter. The cluster is then matched to a reconstructed track. The electron selection is based on the shower shape,
track-cluster matching, and consistency between the cluster energy and the track momentum~\cite{Khachatryan:2015hwa}. Muon candidates are
reconstructed by performing a global fit that requires consistent hit patterns in the tracker and the muon system~\cite{MUOART}.
For both lepton flavors, the impact parameter with respect to the primary vertex is required to be less than 0.5\unit{mm}
in the transverse plane and 1\unit{mm} along the beam direction.

Leptons are required to be isolated from other activity in the event. A measure of lepton isolation is the scalar \pt  sum
 ($\pt^\text{sum}$) of all PF candidates not associated with the lepton within a cone of radius
$\Delta R \equiv\sqrt{\smash[b]{(\Delta\eta)^2+(\Delta\phi)^2}} = 0.3$,
where $\Delta \eta$ and $\Delta \phi$ are the distances between the lepton and the PF candidates
at the primary vertex in $\eta$--$\phi$ space~\cite{TRK-11-001}.
Only charged PF candidates compatible with the primary vertex are included in the sum.
The average contribution of particles from additional $\Pp\Pp$ interactions in the
same or nearby bunch crossings (pileup) is subtracted from $\pt^\text{sum}$. We require
that $\pt^\text{sum}$ be less than 5\GeV.  Typical lepton identification and isolation efficiencies, measured in
samples of $\ensuremath{\cPZ/\Pgg^\star} \to \ell \ell$ events, are approximately 80--85\% (85--90\%) for electrons (muons),
depending on \pt  and $\eta$.

Particle-flow candidates are clustered to form jets using the anti-\kt clustering algorithm~\cite{antikt} with a distance parameter of 0.4, as
implemented in the {\FASTJET} package~\cite{FastJet}.
Only charged PF candidates compatible with the primary vertex are used in the clustering.
The pileup contribution to the jet energy is estimated on an event-by-event basis using
the jet area method described in \cite{cacciari-2008-659} and is subtracted from the overall jet \pt.
Corrections are applied to the energy measurements of jets to account for non-uniform detector response
and are propagated consistently as a correction
to \ptvecmiss~\cite{Chatrchyan:2011ds,Khachatryan:2016kdb}.
The selected lepton can also be reconstructed as a jet, so any jets within $\Delta R = 0.4$ of the
lepton are removed from the list of considered jets.

Selected events are required to contain exactly two jets with $\pt > 30$\GeV and $\abs{\eta} < 2.4$. Both of these jets must
be consistent with containing the decay products of a heavy-flavor (HF) hadron, as identified using the combined secondary
vertex (CSVv2) tagging algorithm~\cite{ref:btag}. Such jets are referred to as b jets.
The CSVv2 algorithm has three main operating points: loose, medium, and tight.
We require both jets to be tagged according to the loose operating point, and at least
one of them to be tagged with the medium operating point.
The efficiency of this algorithm
for jets arising from b quarks with $\pt$ between 30
and 400\GeV is in the range 60--65\% (70--75\%)
for the medium (loose) working point.
The misidentification rate for jets arising from light quarks or gluons is approximately 1\% (10\%) for the medium (loose) working point.

The largest background in this search arises from \ttbar and $\PQt\PW$
events with decays into two-lepton final states
in which one of the leptons is not
reconstructed or identified. In order to reduce these backgrounds, we search for
a second electron or muon
with $\pt > 5\GeV$ and looser identification and isolation requirements, and reject
events where such a lepton is found.
Second leptons are required to satisfy $\pt^\text{sum}/\pt < 0.1$, where $\pt^\text{sum}$
is calculated here with a cone radius of $\Delta R$ = 0.2 for $\pt^{\text{lep}} \le 50$ \GeV, and $\Delta R =
\max(0.05, 10\GeV/\pt^{\text{lep}})$ at higher values of lepton transverse momentum.
We also reject events
with reconstructed hadronically decaying tau leptons with $\pt > 20\GeV$~\cite{Khachatryan:2015dfa},
or isolated tracks with $\pt > 10\GeV$ and opposite electric charge relative to the selected lepton.
For this purpose, a track is considered isolated if $\pt^\text{sum}/\pt < 0.1$ and $\pt^\text{sum} < 6$\GeV, where
$\pt^\text{sum}$ here is constructed with charged PF candidates compatible with the primary vertex, the cone radius is $\Delta R = 0.3$, and \pt is the transverse momentum of the track.

The final two requirements that complete the preselection are $\ETm \geq 125$\GeV and $\mt > 50$\GeV,
where \mt is the transverse mass of
the lepton-\ETm system, defined as
\begin{equation}
 \mt = \sqrt{2 \pt^{\ell} \ETm [1 - \cos(\Delta\phi)]},
\end{equation}
where $\pt^{\ell}$ is the transverse momentum of the lepton
and $\Delta\phi$ is the angle between the
transverse momentum of the lepton and \ptvecmiss.

\subsection{Signal region definition}

\begin{figure}[hbt]
  \centering
        \includegraphics[width=0.45\linewidth]{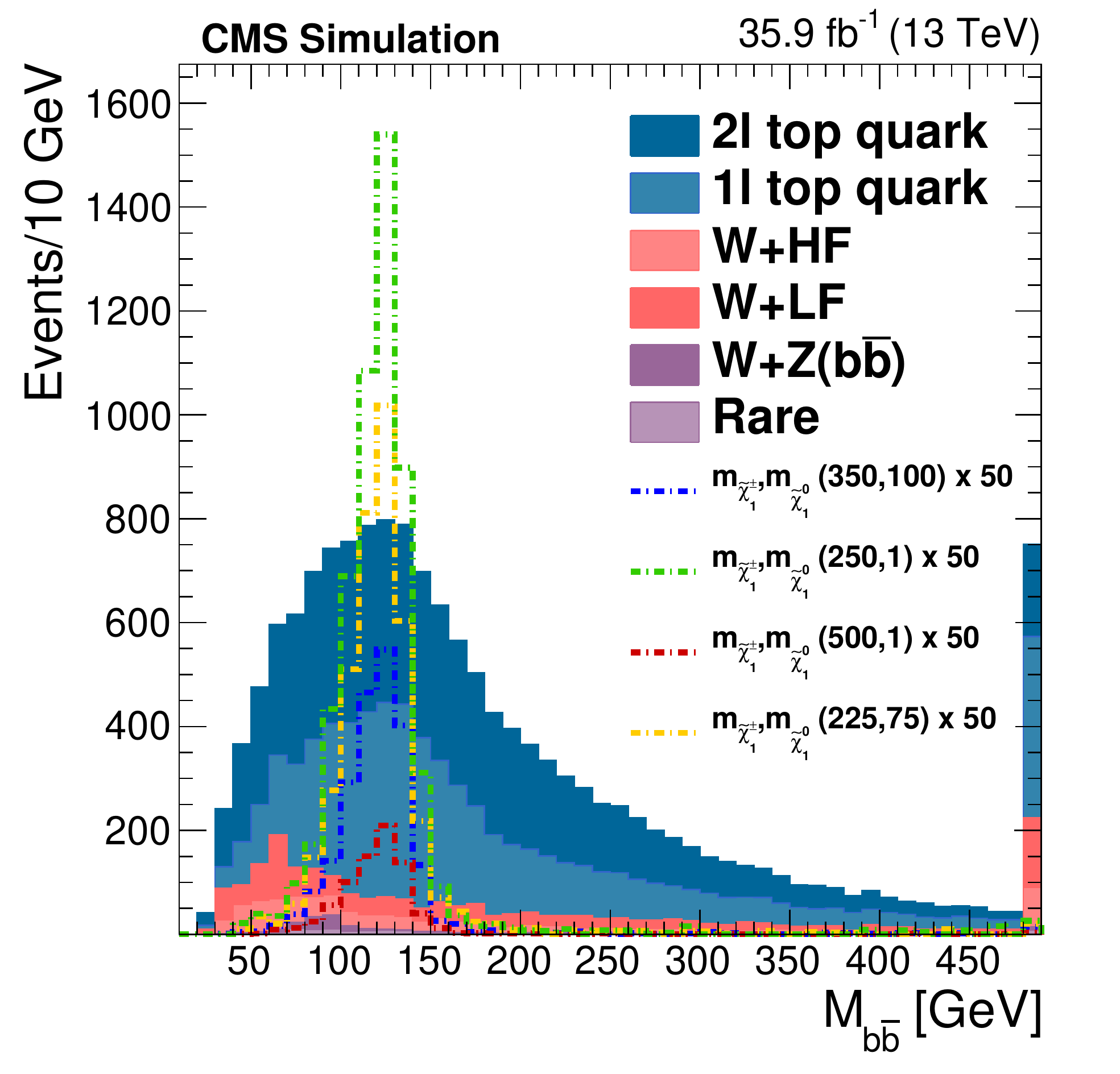}
        \includegraphics[width=0.45\linewidth]{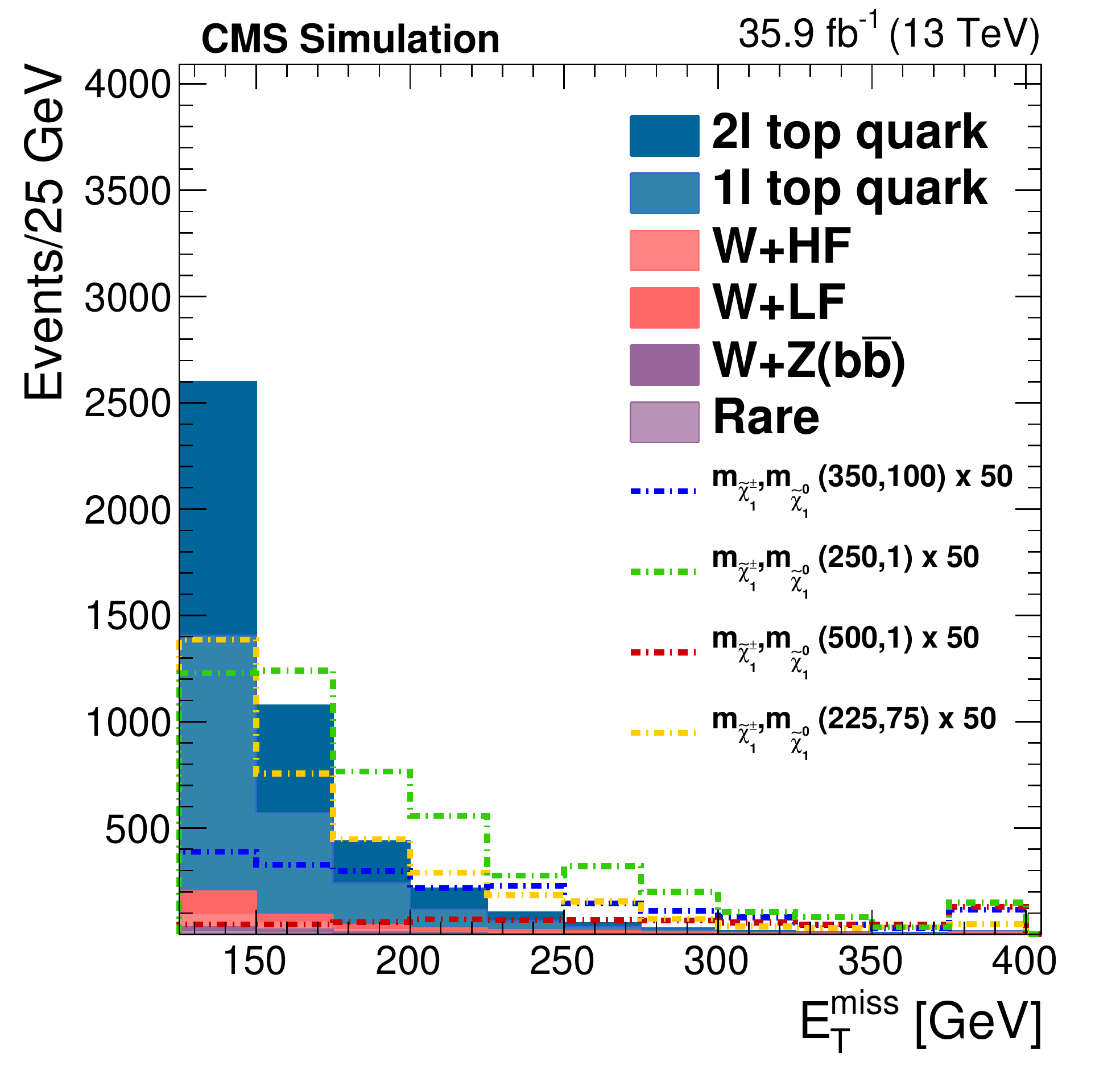}
        \includegraphics[width=0.45\linewidth]{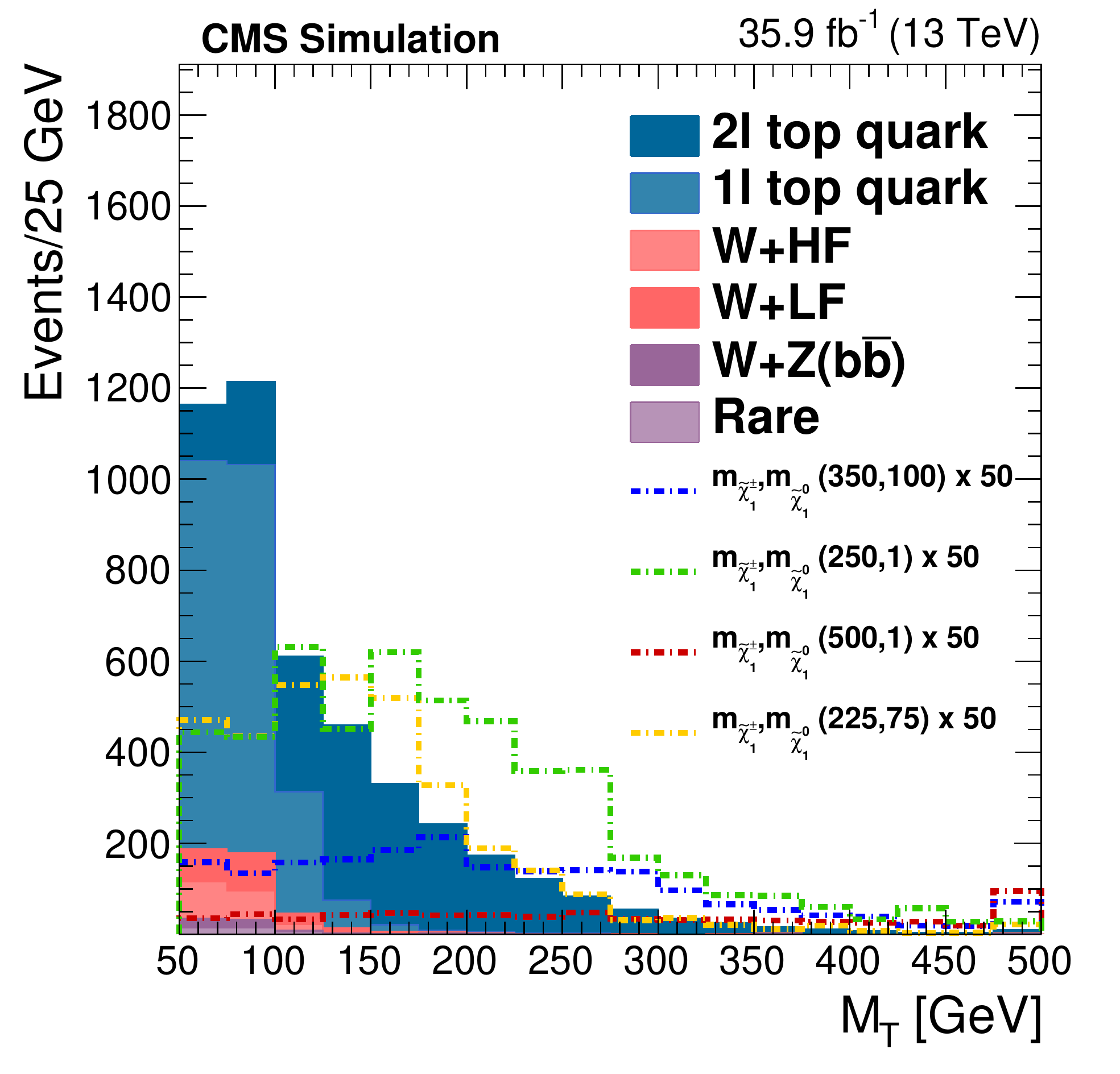}
        \includegraphics[width=0.45\linewidth]{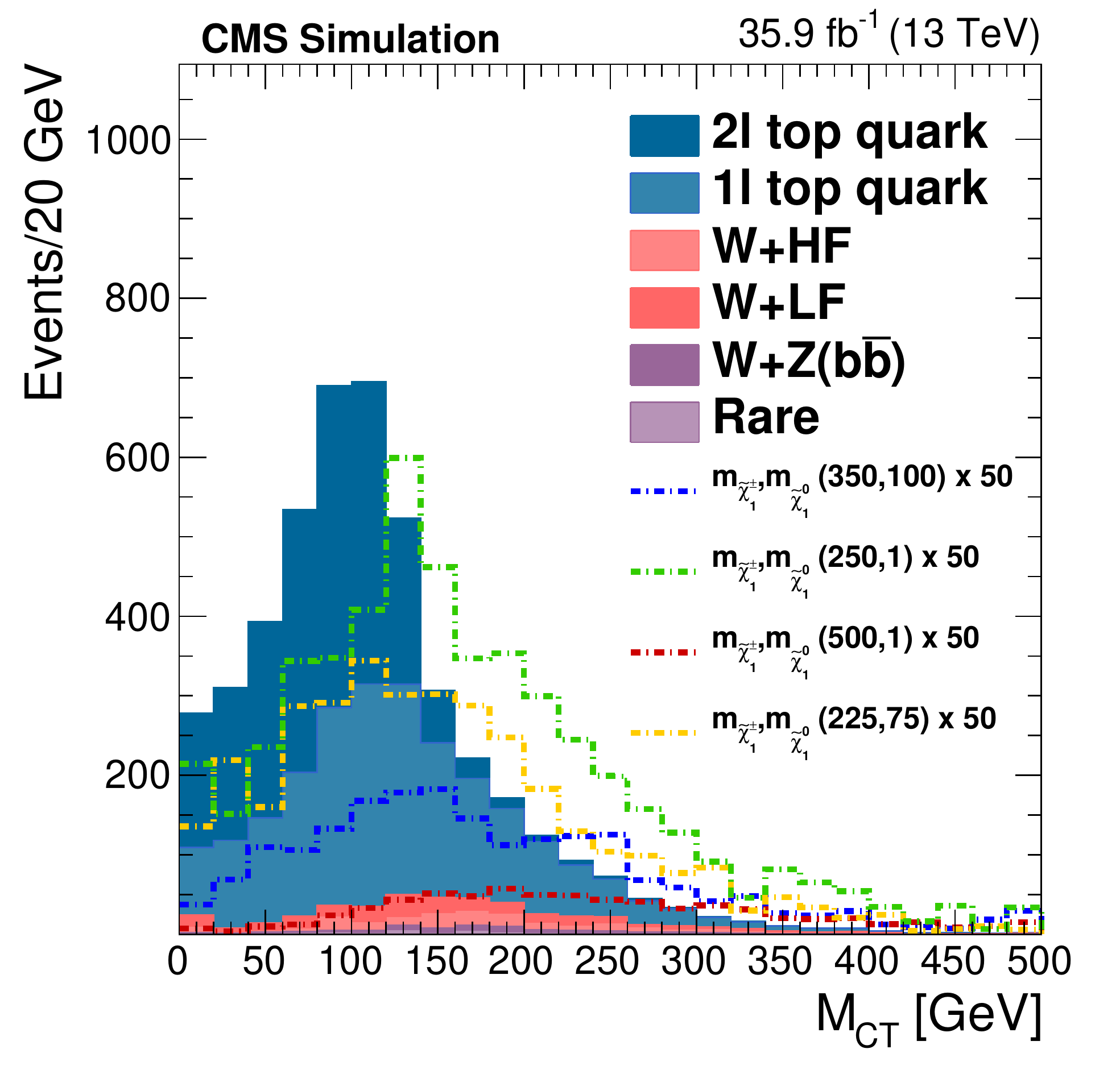}
    \caption{
      Distributions in \mbb (top left), \ETm  (top right), \mt (bottom left), and \mct (bottom right)
      for signal and background events in simulation after the preselection.
      The \ETm, \mt, and \mct distributions are shown after the $90<\mbb<150$\GeV requirement.
      Expected signal distributions are also overlaid as open histograms for various mass points,
      with the signal cross section scaled up by a factor of 50 for display purposes.
      The legend entries for signal give the masses $(m_{\chipmo},
      m_{\lsp})$ in \GeV and the factor by which
      the signal cross section has been scaled.
\label{fig:mt_mct}
}

\end{figure}

The signal regions are defined by additional requirements on the kinematic properties of preselected events.
The invariant mass of the two b jets is required to be in the range $90 \leq \mbb \leq 150$\GeV,
consistent with the Higgs boson mass within the resolution. The \mbb distribution for signal and background processes is shown in
Fig.~\ref{fig:mt_mct} (top left), displaying a clear peak for signal events near the Higgs boson mass.

To suppress single-lepton backgrounds originating from semileptonic
\ttbar, \wjets, and single top quark processes, the preselection requirement
on \mt is tightened to $>$150\GeV.
This is because the \mt\ distribution in these processes with a
single leptonically decaying \PW\ boson has a kinematic endpoint
$\mt < m_{\PW}$, where $m_{\PW}$ is the \PW\ boson mass.  The endpoint can be exceeded by off mass-shell $\PW$ bosons or because of detector resolution effects.
The \mt requirement significantly reduces single-lepton backgrounds, as shown in Fig.~\ref{fig:mt_mct} (bottom left).

In order to further suppress both semileptonic and dileptonic \ttbar
backgrounds,
we utilize the contransverse mass variable,
\mct~\cite{Tovey:2008ui,Polesello:2009rn}:
\begin{equation}
\mct = \sqrt{2 \pt^{\cPqb 1} \pt^{\cPqb 2} [1 + \cos(\Delta \phi_{\cPqb\cPqb})]},
\label{eq:mct}
\end{equation}
where $\pt^{\cPqb 1}$ and $\pt^{\cPqb 2}$ are the transverse momenta of the two jets, and $\Delta \phi_{\cPqb\cPqb}$
is the azimuthal angle between the pair.
As shown in Refs.~\cite{Tovey:2008ui,Polesello:2009rn}, this variable has a kinematic endpoint at $(m^{2}(\delta)-m^{2}(\alpha))/m(\delta)$,
 where $\delta$ is the pair-produced heavy particle and $\alpha$ is the invisible particle produced in the decay of $\delta$.
In the case of \ttbar events, when both jets from b quarks are correctly identified,
the kinematic endpoint corresponds to the top quark mass, while signal events tend to have higher values of \mct.
This is shown in Fig.~\ref{fig:mt_mct} (bottom right). We require $\mct > 170$\GeV.

After all other selections, we define two exclusive bins in \ETm  to enhance sensitivity to signal models
with different mass spectra: $125 \leq \ETm < 200\GeV$ and $\ETm \geq 200\GeV$.
The \ETm distribution is shown in Fig.~\ref{fig:mt_mct} (top right).

\subsection{Signal and background simulation}
Samples of \ttbar, \wjets, and \zjets events, as well as \ttbar production in association with a vector boson,
are generated using the \MGvATNLO 2.2.2~\cite{Alwall:2014hca}
generator at leading order (LO) with the MLM matching
scheme~\cite{Alwall:2007fs},
while
$\PQt\PW$ and single top quark $t$-channel events are generated at next-to-leading-order (NLO) using \POWHEG\ V2~\cite{Nason:2004rx,Frixione:2007vw,Alioli:2010xd}.
A top quark mass of $m_\cPqt=172.5$\GeV, and the NNPDF3.0 LO or NLO~\cite{Ball:2014uwa} parton distribution functions (PDFs) are used in the event generation.
Single top quark $s$-channel production is simulated using \MGvATNLO 2.2.2
at NLO precision with the FxFx matching scheme~\cite{Frederix:2012ps}.
Samples of diboson ($\PW\PW$, $\PW\cPZ$, and $\cPZ\cPZ$) events are generated with either \POWHEG or
\MGvATNLO at NLO precision.
Normalization of the simulated background samples is performed using
the most accurate cross section calculations
available~\cite{Alioli:2009je,Re:2010bp,Alwall:2014hca,Melia:2011tj,Beneke:2011mq,
Cacciari:2011hy,Baernreuther:2012ws,Czakon:2012zr,Czakon:2012pz,Czakon:2013goa,
Gavin:2012sy,Gavin:2010az},
which generally correspond to NLO or next-to-NLO precision.

The chargino-neutralino signal samples
are also generated with \MGvATNLO
at LO precision.  For these samples
we improve on the modeling of initial-state radiation (ISR), which affects the total transverse
momentum ($\pt^\mathrm{ISR}$) of the system of SUSY particles, by
reweighting the
 $\pt^\mathrm{ISR}$ distribution in these events.
This reweighting procedure is based on studies of the \pt  of \cPZ\ bosons~\cite{Chatrchyan:2013xna}.
The reweighting factors range between 1.18 at  $\pt^\mathrm{ISR} = 125$\GeV
 and  0.78 for $\pt^\mathrm{ISR} > 600$\GeV.
We take the deviation from 1.0 as the systematic uncertainty in the reweighting
procedure.

Parton showering and fragmentation in all of these samples are performed using \PYTHIA V8.212~\cite{Sjostrand:2007gs}
with the CUETP8M1 tune \cite{Khachatryan:2015pea}.
For both signal and background events, additional simultaneous proton-proton interactions (pileup)
are generated with \PYTHIA and superimposed on the hard collisions.
The response of the CMS detector for SM background samples is simulated using \GEANTfour-based model~\cite{Geant}, while that for new
physics signals is performed using the CMS fast simulation package~\cite{fastsim}.
All simulated events are processed with the same chain of reconstruction programs as that used for collision data.

Small differences between the \PQb~tagging efficiencies measured in data and
simulation
are corrected using data-to-simulation scale factors.
Corrections are also applied to account for differences between lepton selection efficiencies (trigger, reconstruction, identification,
and isolation) in data and simulation.

\section{Backgrounds}
\label{sec:bg}

The backgrounds for this search are classified into six categories.
The first and most important category, referred to as dilepton top quark events, consists mainly of events
from top quark pair production with both quarks decaying leptonically,
but also including contributions from the associated production of a
single top quark with a \PW\ boson, both of which decay leptonically.
The second to fifth categories include processes with a single leptonically decaying \PW\ boson.
Events with a single \PW\ are divided into two categories: those with b
jets (\whf, for ``heavy flavor'') and those without (\wl, for ``light flavor'').
A separate category comprises $\PW\cPZ$ events
in which the \cPZ\ boson decays to \bbbar (\wzbb).
Events with one leptonically decaying top quark, either from
\ttbar or from single top quark $t$- or $s$-channel production, are
included
in the fifth category (``single-lepton top quark'').
Finally, other SM processes contribute a small amount to the expected yield in the signal region
and are grouped together in the ``rare'' category.  This includes
events from
\zjets, \PW\PW, \PW\Z (except the decays described above), $\Z\Z$, triboson,
\ttw, \ttz, and \whbb\ processes.

All background processes are modeled using MC simulation.
Three data control regions (CRs) are defined by inverting the signal region
selection requirements, as summarized in Table~\ref{tab:crsel}.
The CRs are defined at both preselection and signal region selection levels.
The CRs at the preselection level are defined with looser cuts in order to check the modeling of key discriminant variables.
The CRs after the signal region level selections are used to validate the modeling of the main backgrounds and to
assign systematic uncertainties in the background predictions.
The regions \crmbb and \crzb are split into two bins in \ETm  to mirror the signal region selection.
The expected signal contribution in any of the CRs is always less than 1\% of the total SM yields,
and typically much smaller.

\begin{table}[htb]
\centering
\topcaption{\label{tab:crsel}
Event selections in signal and control regions. The region \crdl\ is only used at the preselection level.}
\resizebox{\textwidth}{!}{
\begin{tabular}{lcccc}

Selection & Signal regions & \crdl & \crzb & \crmbb\ \\
\hline
N(leptons)                     & $=$1               & $=$1 or 2          & $=$1               & $=$1                     \\
Isolated track veto            & $\checkmark$        & inverted if 1$\ell$ & $\checkmark$        & $\checkmark$              \\
Tau candidate veto             & $\checkmark$        & inverted if 1$\ell$ & $\checkmark$        & $\checkmark$              \\
Number of b tags               & $=$2               & $=$2               & $=$0               & $=$2                     \\
\hline
 & &Preselection level &&\\
\hline
\mbb  [\GeVns{}]                          & \NA   & \NA                   &  $\in$[90,150]& $\notin$[90,150]    \\
\MET [\GeVns{}]                 & $\geq$125 &   $\geq$125 &      $\geq$125      & $\geq$125 \\
\mt   [\GeVns{}]        & $>$50       & $ >$150        &  $>$50      & $>$150     \\
\mct [\GeVns{}]   & \NA       & \NA                   &  $>$170      & \NA    \\
\hline
&&Signal region selection level &&\\
\hline
\mbb  [\GeVns{}]                         & $\in$[90,150]   & \multirow{4}{*}{not used}                  & $\in$[90,150]& $\notin$[90,150]\\
\ETm [\GeVns{}]                         & $[125,200), \geq 200$&  & [125,200), $\geq$200& $[125,200), \geq 200$\\
\mt   [\GeVns{}]                         & $>$150       &         &  $ >$150      & $ >$150    \\
\mct [\GeVns{}]                         & $>$170       &                    &  $ >$170      & $ >$170    \\
\hline
\end{tabular}
}
\end{table}

The dilepton top quark background can be isolated in the \crdl control region by selecting dilepton events.
In addition to a lepton passing the analysis selections, events must contain one of the following:
a second electron or muon, an isolated track candidate, or a tau lepton candidate.
The latter categories are included to accept hadronically decaying tau leptons.
If all the kinematic selections used for the signal regions are applied,
the number of events in \crdl is too low to validate the modeling of
the dilepton top quark background.
Therefore, this CR is used primarily to validate the modeling of \mbb.

Since the signal produces a resonant peak in the \mbb distribution, the
requirement on \mbb is inverted to define the background-dominated control region
\crmbb, which includes a mixture of all backgrounds in proportions similar to those in the signal region.
Consequently, this control region is dominated by the dilepton top quark background
and is used to validate the modeling of these processes in the kinematic tails of the
\ETm, \mt, and \mct distributions.

The \crzb region is designed to study the \wl background.
It is used to validate the modeling of the kinematic tails
in \ETm, \mt, and \mct for \wjets processes.
In this region, the dijet mass \mjj computed from the two selected jets is used in place of \mbb.

The background estimation and the associated uncertainties are
described
in the following sections.

\subsection{Dilepton top quark backgrounds}
\label{sec:diltop}

\begin{figure}[hbt]
  \centering

        \includegraphics[width=0.49\linewidth]{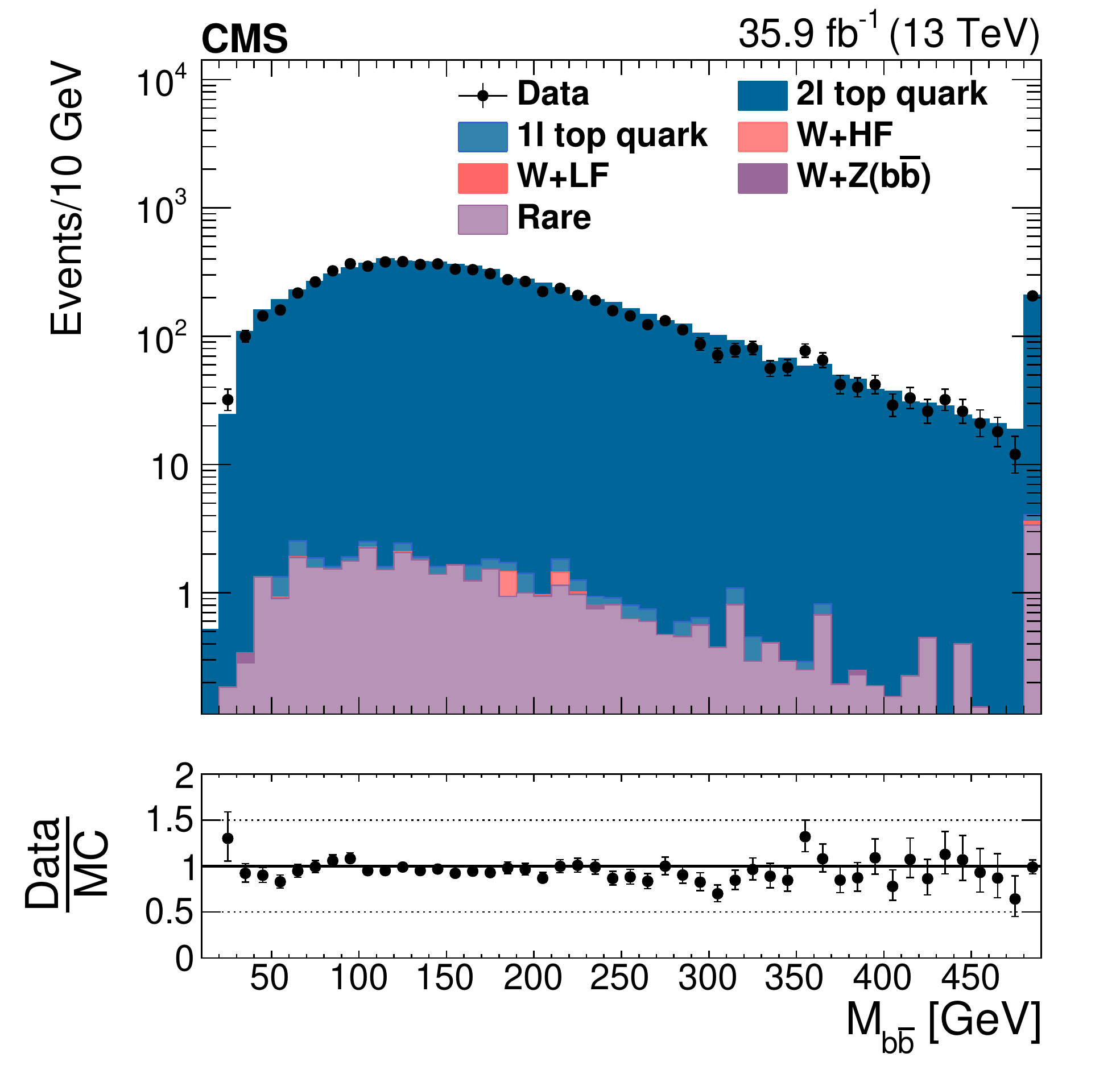}
        \includegraphics[width=0.49\linewidth]{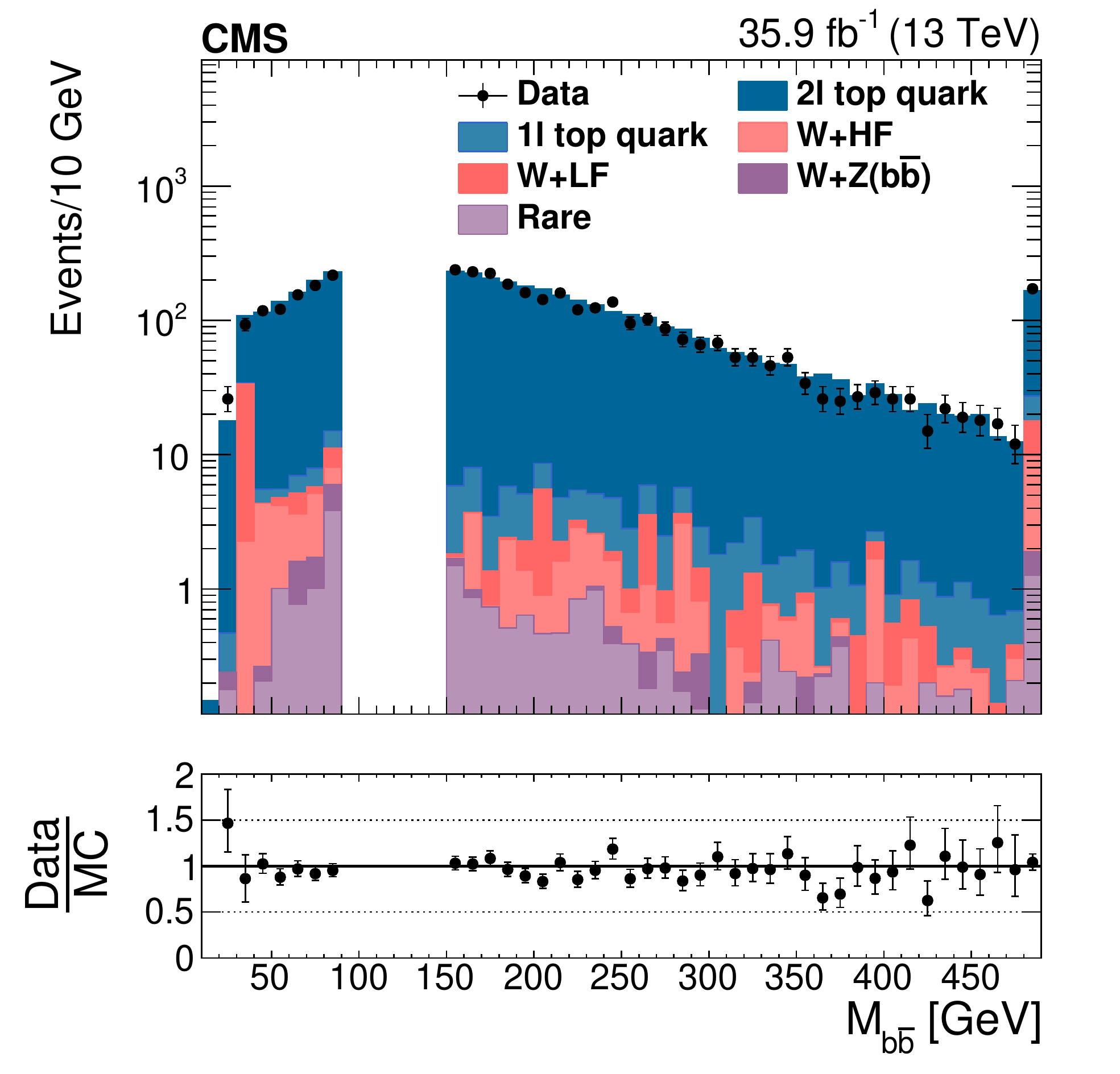}
    \caption{
      (left) Distribution in \mbb in \crdl\ after the preselection level cuts defined in Table~\ref{tab:crsel}, comparing data to MC simulation.
      (right) Distribution in \mbb in \crmbb\ after preselection level cuts defined in Table~\ref{tab:crsel}.
      The signal region range of $90 \leq \mbb \leq 150$\GeV has been removed from the plot.
\label{fig:crs_mbb}
}

\end{figure}

The dilepton top quark process contributes to the event sample in the signal region when the second lepton is not reconstructed or identified.
Due to the presence of two neutrinos, these events tend to have higher \ETm
than the single-lepton backgrounds, and their \mt distribution is
not bounded by the \PW\ boson mass.
However, as mentioned above, the \mct requirement significantly suppresses dilepton top quark events.
The modeling of this background is validated in two steps.
First, the modeling of the \mbb distribution is validated in \crdl;
second, the modeling in the kinematic tails of the \ETm, \mt, and \mct distributions is validated in \crmbb.
Distributions of \mbb in \crdl and \crmbb, after the preselection level cuts defined in Table~\ref{tab:crsel}, are displayed in
Fig.~\ref{fig:crs_mbb}~(left)
and Fig.~\ref{fig:crs_mbb}~(right), respectively.

In \crdl, we observe agreement between data and MC, validating the modeling of the \mbb distribution.
We then use \crmbb\ at the signal region selection level to derive a scale factor
for the dilepton top quark background separately in each of the analysis \ETm bins.
All other background components are subtracted from the observed data yields,
and the result is compared to the dilepton top quark MC prediction.
Agreement is observed in the higher \ETm  bin within statistical uncertainties.
For the lower \ETm bin, we find fewer events in data than predicted,
and we derive a scale factor of 0.72 for the dilepton top quark background in this bin.
From the statistical precision of the data, we assign a systematic uncertainty of 30\% in the prediction for both bins.
This accounts for any effects that could impact the modeling of this background in simulation,
including generator assumptions on factorization and renormalization scales, and PDFs, as
well as experimental uncertainties in the jet energy scale, the
lepton identification and isolation, trigger, and b tagging efficiencies.

\subsection{W boson backgrounds}
\label{sec:wjets}

The \mt\ requirement ($>$150\GeV) effectively suppresses the contribution from \wjets events.
However, as discussed above, events from \wjets can still enter the \mt tail due to
off-shell \PW\ production or \ETm  resolution effects.
The control region \crzb consists mostly of \wl events and is therefore used to validate the modeling of \wjets
in the tails of all kinematic variables such as \mt.

\begin{figure}[hbt]
  \centering
        \includegraphics[width=0.5\linewidth]{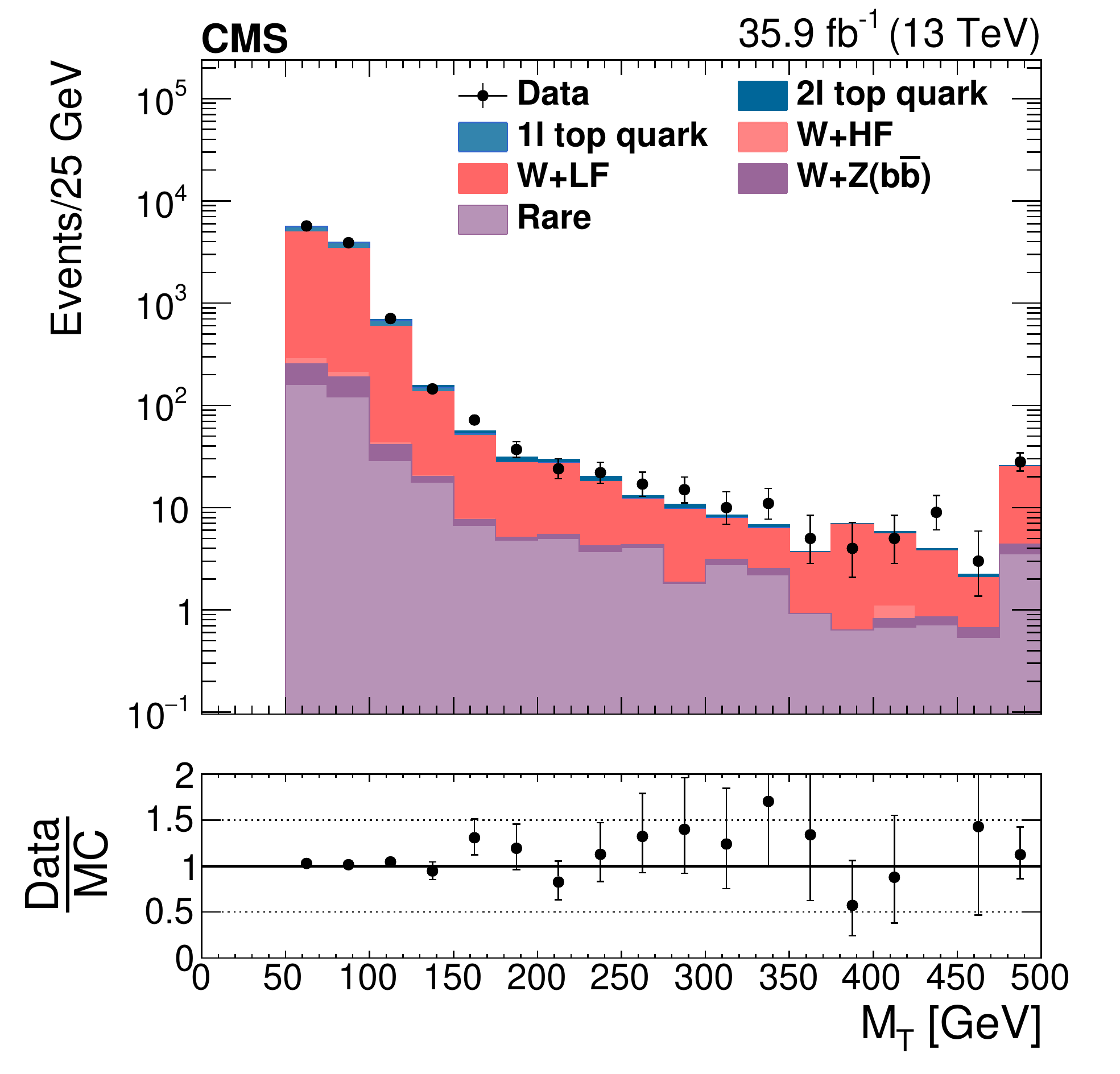}
    \caption{
Distribution in \mt in \crzb after the preselection level cuts defined in Table~\ref{tab:crsel}.
\label{fig:crzb_mt}
}

\end{figure}

Figure~\ref{fig:crzb_mt} shows the \mt\ distributions of data and
simulated events in \crzb after the preselection requirements.  The data and simulation agree within uncertainties.
The observed yields in data are then compared with MC predictions
after applying all the kinematic requirements at signal region selection level defined in Table~\ref{tab:crsel}.
We find agreement within statistical uncertainties.
Based on the statistical precision of the data, we assign a 10\% systematic uncertainty in the \wjets prediction.
This procedure directly tests the \wjets background prediction in the kinematic phase space of the signal region,
including experimental uncertainties in the jet energy scale, in the efficiencies for trigger,
lepton identification and isolation.  It also accounts for most generator assumptions.
Additional uncertainties for effects not tested by this procedure are discussed below.

For the \wl background, the uncertainty due to the b tagging requirements
is evaluated by varying the b tagging efficiencies
within their measured uncertainties.  The uncertainty in the yield in
the signal regions is 1\%.

For the \whf background the effects contributing to the kinematic tails
are similar
to those in \wl.
In this case the tail of the \mt distribution receives
contributions from off-shell \PW\ boson production and \ETm  resolution
effects, but also
from neutrinos in semileptonic decays within the b jets.  Since this
last effect
is accounted for in the event generation,
we do not apply any additional correction or uncertainty for kinematic tail modeling beyond the one
derived above in \crzb.

The most uncertain aspect of the prediction for the \whf background is the estimate of its cross
section relative to the \wl process.  We assign a 50\% uncertainty to the
normalization of this
background~\cite{Sirunyan:2016jpr}.
This uncertainty is validated by comparing data to simulation in a
CR with $60 < \mt < 120$\GeV and with one or two jets, where
the dominant contribution to the event sample is from \wjets.
We find that the 50\% uncertainty
conservatively covers differences between data and simulation as a function of the
number of b jets.  Finally, the uncertainty in this prediction due to the uncertainty
in the b tagging efficiency is also evaluated and found to be 5\%.

The effects discussed above also contribute to the tail of the \mt distribution in \wzbb events.
As a result, the tail modeling systematic uncertainty for this background is taken
to be the same as those evaluated in \crzb.
An additional uncertainty of 12\% is applied to the normalization of the \wzbb background,
based on the CMS cross section measurement of inclusive $\PW\cPZ$ production at 13\TeV~\cite{cmsWZ13TeV}.
A unique aspect of the \wzbb background is that \mbb peaks at the \cPZ\ boson mass, at the lower edge of the \mbb selection
used in this analysis.  Uncertainties in the jet energy scale can therefore strongly impact the
prediction of this background.  By varying the jet energy scale within
its uncertainty, we derive an uncertainty of
27\% in the \wzbb background prediction.  While this uncertainty is large, the absolute magnitude of this background remains very small in the signal region.
Finally, the uncertainty in the background prediction for this process due to the uncertainty
in the b tagging efficiency is 2\%.

\subsection{Other backgrounds}
\label{sec:rare}

The single-lepton top quark backgrounds are highly suppressed by several of the selections applied in this analysis.
Since these contain exactly one leptonically decaying \PW\ boson, the \mt requirement is an effective discriminant against them.
Requiring exactly two jets also suppresses the \ttsl background, which typically has four jets in the final state.
As a result, this background comprises a small fraction of the expected SM prediction in the signal region.

Isolating the single-lepton top quark background in a region kinematically similar to the signal region is difficult
since dilepton top quark events tend to dominate when requiring large \mt.
The main source of uncertainty in the prediction of this backgrounds is the
modeling of the \ETm  resolution, which was found to be well modeled in the study of \crzb.

Additional studies of \ETm  resolution are performed using \gjets\ events following the method used in Ref.~\cite{Sirunyan:2016jpr}.
The resolution in data is found to be up to 20\% worse than in simulation, leading to higher single-lepton top quark
yields than expected from simulation. However, the impact of this effect on the total background
prediction is negligible.   Due to the difficulties in defining a dedicated control
region for this process, we assign a conservative uncertainty of 100\% to the
single lepton top quark background prediction.

The ``rare'' backgrounds contribute less than 15\% of the expected yield in the signal region.
We apply an uncertainty of 50\% on the event yields from these processes.

\section{Results}

\begin{figure}[hbt]
  \centering
        \includegraphics[width=0.48\linewidth]{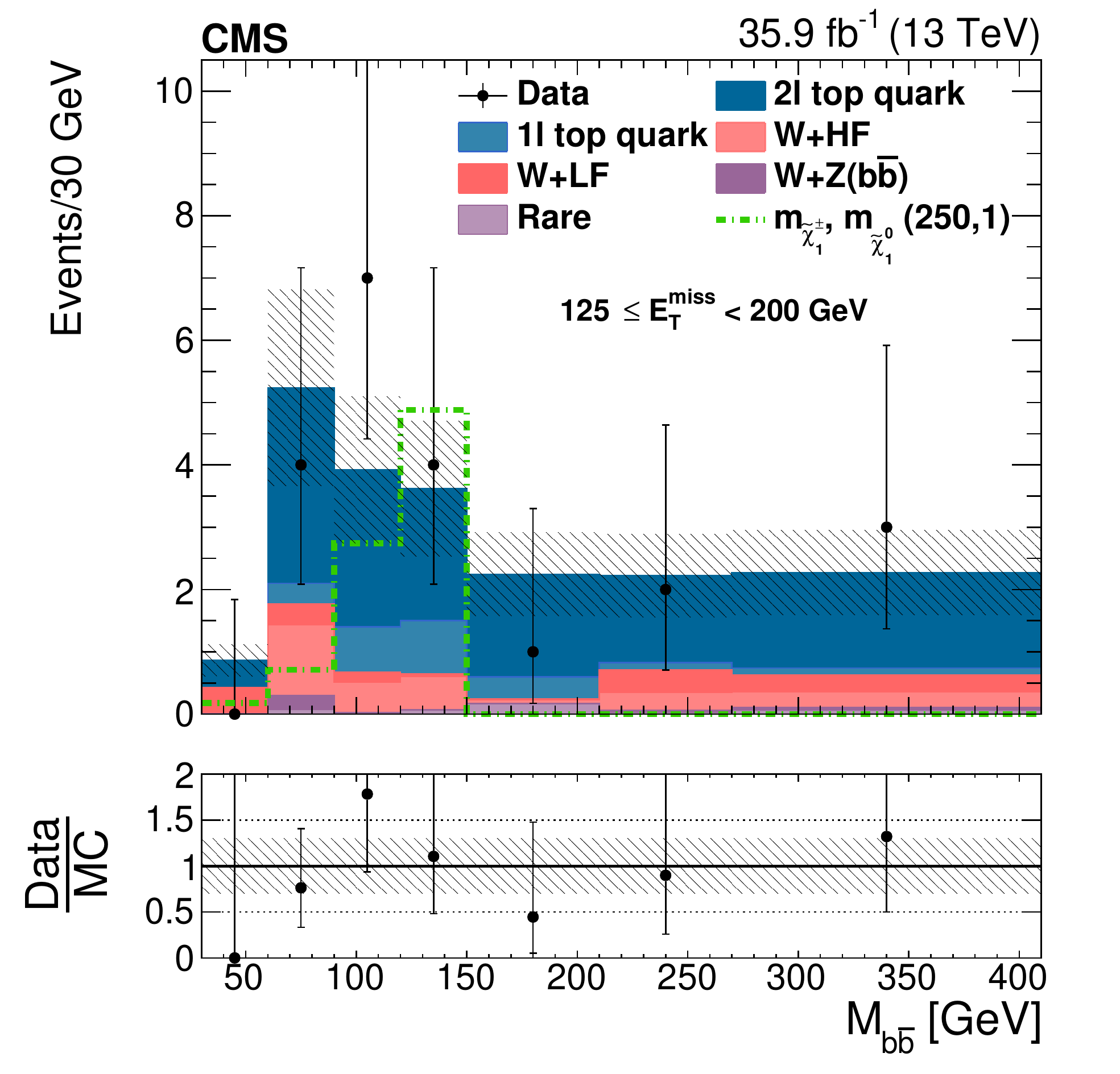}
        \includegraphics[width=0.48\linewidth]{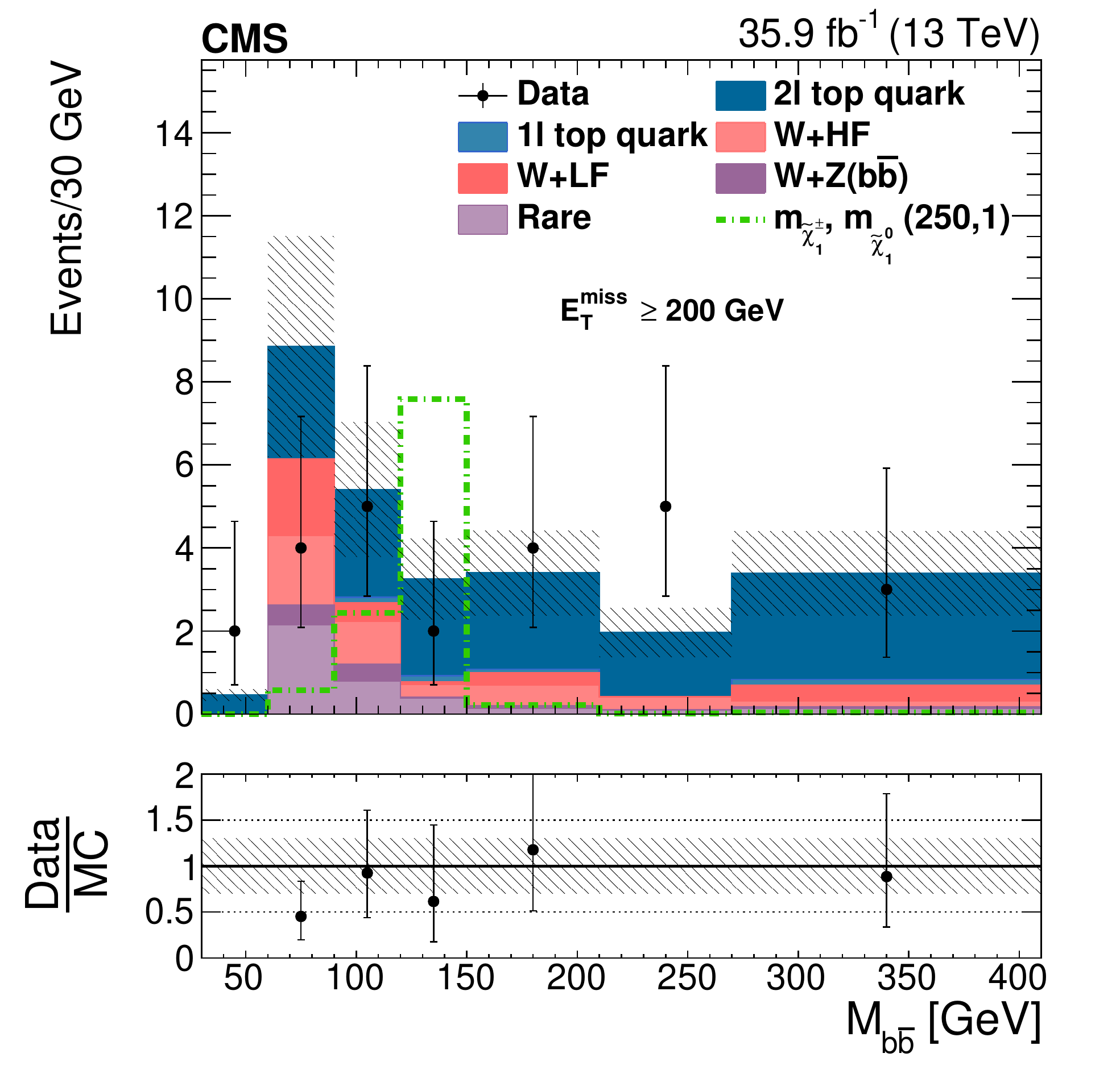}
    \caption{
      Distributions in \mbb after all signal region kinematic
      requirements for the two exclusive \ETm  bins (left: $125 \leq \ETm < 200$\GeV,
      right: $\ETm \geq 200$\GeV).
      The signal region is $90 \leq \mbb \leq 150$\GeV.
      The hatched band shows the total uncertainty in the background prediction,
      including statistical and systematic components.
      The expected signal distribution for a reference SUSY model is overlaid as an open histogram, and
      the legend (on the last line) gives the masses as $(m_{\chipmo}, m_{\lsp})$ in \GeV.
\label{fig:sr_mbb}
}

\end{figure}

Figure~\ref{fig:sr_mbb} shows the distributions of \mbb in data compared with
the SM background prediction
after all signal region requirements except the \mbb selection. No significant deviations from the predictions are observed.
Table~\ref{tab:yields36fb} shows the expected SM background yields in the signal region compared to the observation,
as well as predicted yields for several signal models with the masses $(m_{\chipmo}, m_{\lsp})$ indicated in~\GeV.
The correlation coefficient for the background prediction between the two bins is 0.61.
The correlation is incorporated in the likelihood model described below for the interpretation of the results,
and it can be used to reinterpret these results in other signal models~\cite{Collaboration:2242860}.

\begin{table}
\centering
  \topcaption{ Expected and observed event yields in the signal regions.
    The uncertainties shown include both statistical and
  systematic sources.
The correlation coefficient for the background prediction between the two bins is 0.61.
Predicted yields are shown also for several signal models with the masses $(m_{\chipmo}, m_{\lsp})$ indicated in \GeV
and with statistical-only uncertainties. \label{tab:yields36fb}}
\begin{tabular}{lcc}

& $125 \leq\ETm< 200$\GeV &$\ETm \geq 200$\GeV\\
\hline
Dilepton top quark&$4.6\pm1.5$&$4.9\pm1.7$\\
\wl&$0.2\pm0.1$&$0.5\pm0.4$\\
\whf&$1.0\pm0.9$&$1.3\pm1.0$\\
\wzbb&$0.1\pm0.1$&$0.4\pm0.2$\\
Single-lepton top quark&$1.6\pm1.6$&$0.3\pm0.4$\\
Rare&$0.0^{+0.2}_{-0.0}$&$1.2\pm0.7$\\[1.5ex]

Total SM background &$7.5\pm2.5$&$8.7\pm2.2$\\[1.5ex]

Data&11&7\\[1.5ex]
$\chipmo\lsp$ (225,75)&$2.4\pm0.4$&$2.3\pm0.4$\\
$\chipmo\lsp$ (250,1)&$7.6\pm1.0$&$10.0\pm1.2$\\
$\chipmo\lsp$ (500,1)&$0.9\pm0.1$&$6.3\pm0.2$\\
$\chipmo\lsp$ (500,125)&$1.0\pm0.1$&$5.5\pm0.2$\\
$\chipmo\lsp$ (350,100)&$2.7\pm0.3$&$8.0\pm0.5$\\
\hline
\end{tabular}

\end{table}
\section{Interpretation}

The results of this analysis are interpreted in the context of the simplified SUSY model
depicted in Fig.~\ref{fig:tchiwh}, $\chipmo\chitn\to\PW^{\pm}\PH\lsp\lsp$.
The $\chipmo$ and $\chitn$ are assumed to have the same mass,
and the branching fractions for the decays listed above are taken to be 100\%.
The \PW\ and Higgs bosons are taken to decay according to their SM branching fractions.
Cross section limits as a function of the SUSY particle masses are set using a modified frequentist approach,
employing the CL$_{\mathrm{s}}$ criterion and an asymptotic formulation~\cite{Read:2002hq,Junk:1999kv,Cowan:2010js, ATL-PHYS-PUB-2011-011}.
Both signal regions are considered simultaneously in setting limits.
The ``expected'' limit is that under the background-only hypothesis,
while the ``observed'' limit reflects the data yields in the signal regions.
The production cross sections are computed at NLO plus next-to-leading-log (NLL) precision
in a limit of mass-degenerate wino $\chipmo$ and $\chitn$, light bino $\lsp$,
and with all the other sparticles assumed to be heavy and decoupled~\cite{Fuks:2012qx,Fuks:2013vua}.
The uncertainty in the cross section calculation includes variations
of factorization and renormalization scales, and of the PDFs.

The systematic uncertainties in the signal yield are summarized in Table~\ref{tab:sig_systs}.
The signal models with the largest acceptance uncertainties are those with $\Delta m = m_{\chitn} - m_{\lsp} \simeq m_{\PH}$.
For these models, the kinematic properties of the events are most similar to those from
SM backgrounds, and as a result, the acceptance is smaller than
for models with larger $\Delta m$.
For these models with compressed mass spectra,
the largest uncertainties in the signal yields arise from the jet energy scale (up to 40\%),
\ETm  resolution in fast simulation (up to 50\%), and limited size of MC samples (up to 60\%).
These uncertainties reach their maximal values only for models where the acceptance of this
analysis is very small and the sensitivity is similarly small.
For models with large $\Delta m$, where this analysis has the best sensitivity,
these uncertainties typically amount to only a few percent.
Other experimental and theoretical uncertainties are also considered and lead to small
changes in the expected yields.  These include effects from the renormalization and factorization
scales assumed in the generator on the signal acceptance, the
b tagging efficiency, the lepton reconstruction, identification, and isolation efficiency,
the trigger efficiency, and the modeling of pileup.
Finally, the uncertainty in the integrated luminosity is 2.5\%~\cite{CMS-PAS-LUM-17-001}.

\begin{table}[hbt]
  \topcaption{Sources of systematic uncertainty in the estimated signal yield, along with their typical values.
    The ranges represent variation across the signal masses probed.
    \label{tab:sig_systs}}
\centering
\begin{tabular}{lc}
Source & Typical range of values [\%] \\
\hline
Integrated luminosity                     & 2.5   \\
Size of MC samples                        & 2--60  \\
Pileup                                    & 1--5       \\
Renormalization and factorization scales  & 1--3       \\
ISR modeling                              & 1--5   \\
b tagging efficiency                      & 2--8   \\
Lepton efficiency                         & 2--5  \\
Trigger efficiency                        & 1--5   \\
Jet energy scale                          & 1--40       \\
Fastsim \ETm  resolution                  & 1--50       \\
\hline
\end{tabular}
\end{table}

\begin{figure}[hbt]
  \centering
        \includegraphics[width=0.44\textwidth]{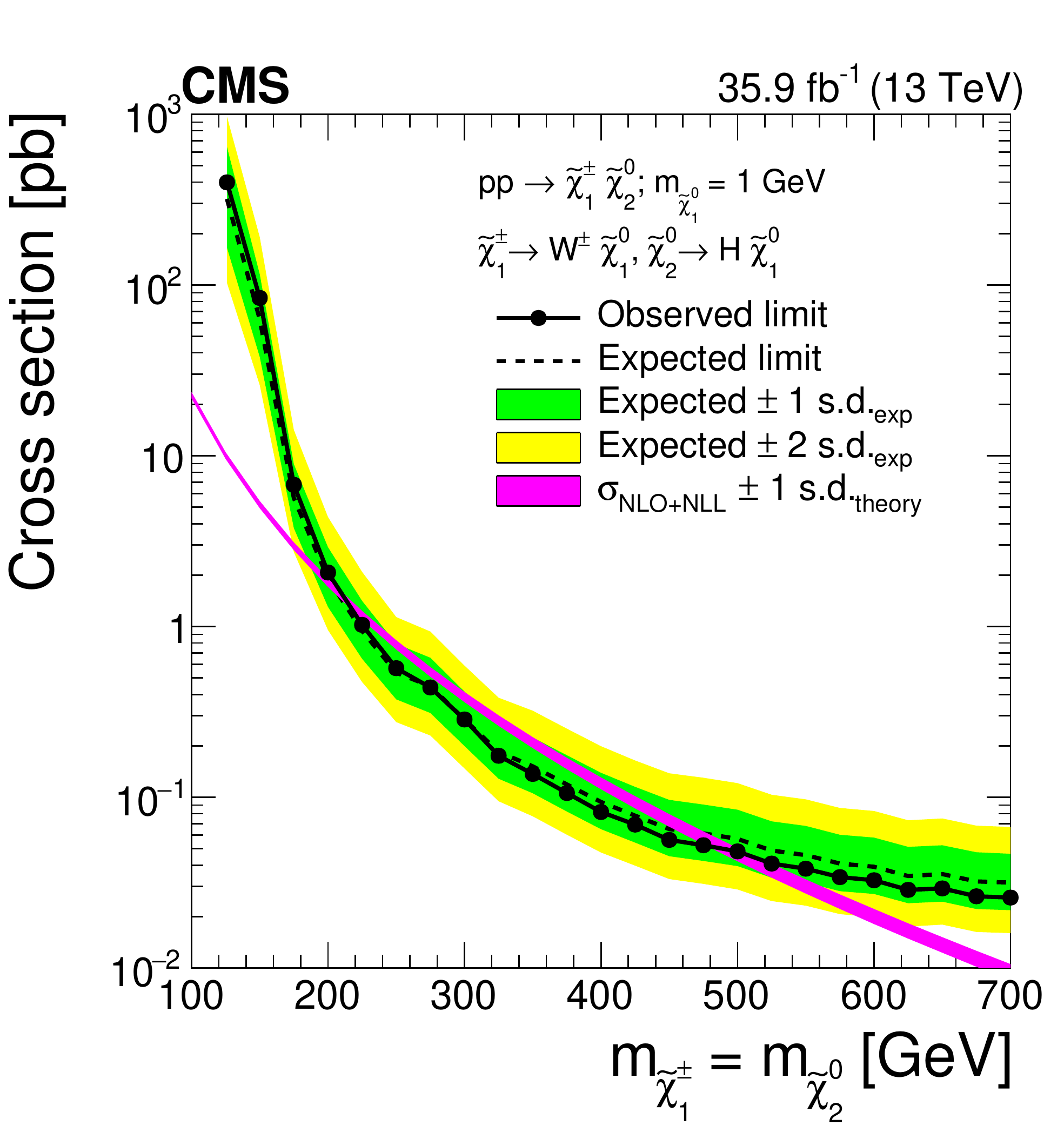}
         \includegraphics[width=0.5\textwidth]{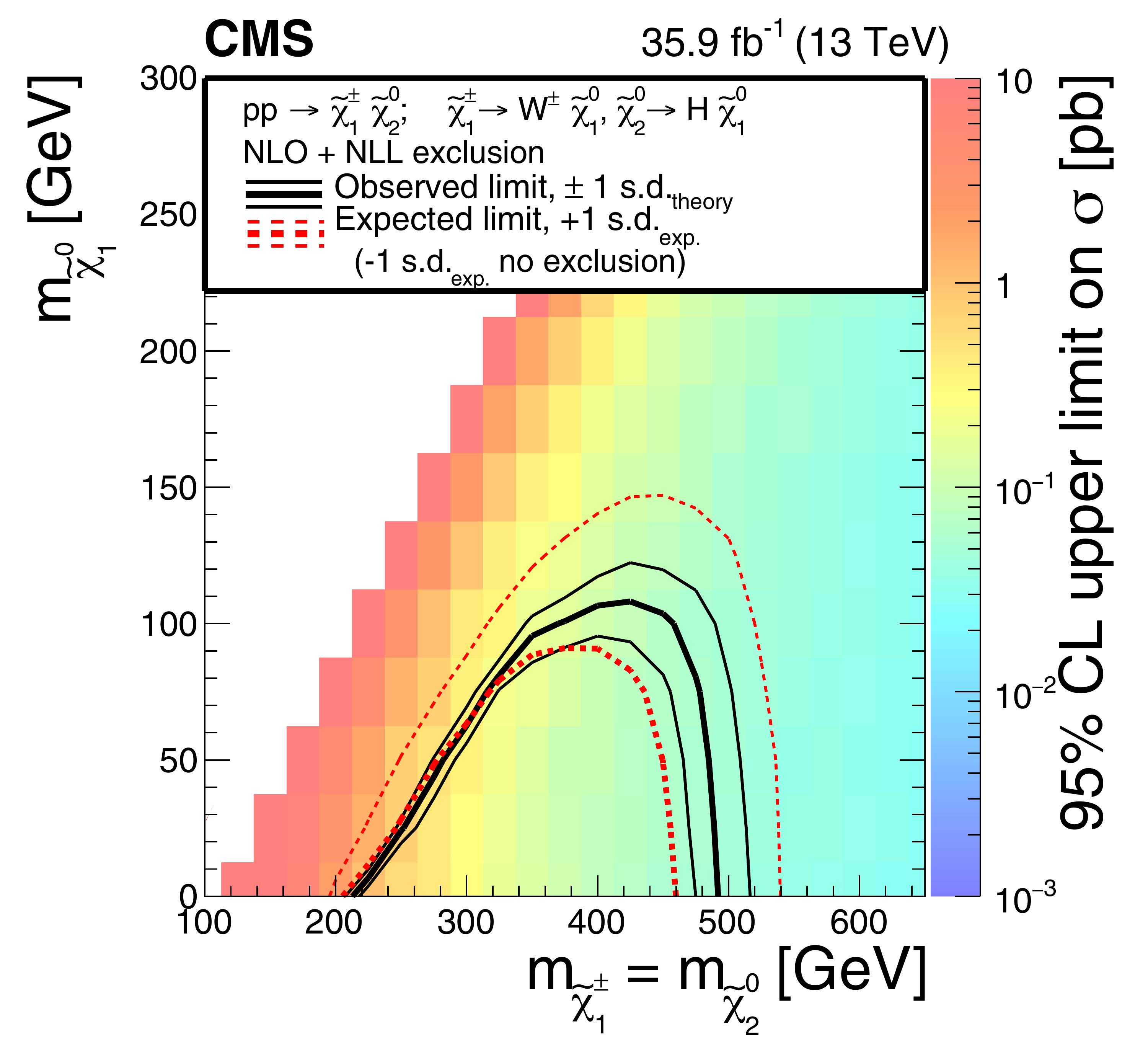}
    \caption{
      (left) Cross section exclusion limits at the 95\% CL are shown for $\chipmo\chitn\to\PW^{\pm}\PH\lsp\lsp$
      as a function of $m_{\chipmo}$, assuming $m_{\lsp} = 1$\GeV.
      The solid black line and points represent the observed exclusion.  The dashed black line represents
      the expected exclusion, while the green and yellow bands indicate the
      $\pm$1 and 2 standard deviation (s.d.) uncertainties in the expected limit.
      The magenta line shows the theoretical cross section with its uncertainty.
(right) Exclusion limits at the 95\% CL  in the plane of $m_{\chipmo}$ and $m_{\lsp}$.
      The area below the thick black (dashed red) curve represents the observed (expected) exclusion region.
      The thin dashed red line indicates the +1 s.d.$_{\text{exp.}}$ experimental uncertainty.
      The -1 s.d.$_{\text{exp.}}$ line does not appear as no mass points would be excluded in that case.
      The thin black lines show the effect of the theoretical
      uncertainties ($\pm$1 s.d.$_{\text{theory}}$) on the signal cross section.
\label{fig:limits}
}

\end{figure}

Figure~\ref{fig:limits} shows the expected and observed 95\% confidence level (CL) exclusion limits
for $\chipmo\chitn\to\PW^{\pm}\PH\lsp\lsp$ as a function of $m_{\chipmo}$ assuming $m_{\lsp} = 1$\GeV (left)
and then in the two-dimensional plane of $m_{\chipmo}$ and $m_{\lsp}$
(right).
This search excludes $m_{\chipmo}$ values between 220 and 490\GeV when $m_{\lsp} = 1$\GeV,
and $m_{\lsp}$ values up to 110\GeV when $m_{\chipmo}$ is around 450\GeV.

\section{Summary}

A search is performed for beyond the standard model physics in events with a leptonically decaying $\PW$ boson,
a Higgs boson decaying to a \bbbar pair, and large transverse momentum imbalance.
The search uses proton-proton collision data recorded by the CMS experiment in 2016 at $\sqrt{s} = 13$\TeV,
corresponding to an integrated luminosity of \fullLumi.
The event yields observed in data are consistent with the estimated standard model backgrounds.
The results are used to set cross section limits on chargino-neutralino production in a simplified supersymmetric model
with degenerate masses for $\chipmo$ and $\chitn$ and
with the decays $\chipmo\to\PW^{\pm}\lsp$ and $\chitn\to\PH\lsp$.
Values of $m_{\chipmo}$ between 220 and 490\GeV are excluded at 95\% confidence level by this search when the $\lsp$ is massless,
and values of $m_{\lsp}$ are excluded up to 110\GeV for $m_{\chipmo} \approx  450$\GeV.
These results significantly extend the previous best limits, by up to 270\GeV in $m_{\chipmo}$ and up to 90\GeV in $m_{\lsp}$.

\begin{acknowledgments}
We congratulate our colleagues in the CERN accelerator departments for the excellent performance of the LHC and thank the technical and administrative staffs at CERN and at other CMS institutes for their contributions to the success of the CMS effort. In addition, we gratefully acknowledge the computing centers and personnel of the Worldwide LHC Computing Grid for delivering so effectively the computing infrastructure essential to our analyses. Finally, we acknowledge the enduring support for the construction and operation of the LHC and the CMS detector provided by the following funding agencies: BMWFW and FWF (Austria); FNRS and FWO (Belgium); CNPq, CAPES, FAPERJ, and FAPESP (Brazil); MES (Bulgaria); CERN; CAS, MoST, and NSFC (China); COLCIENCIAS (Colombia); MSES and CSF (Croatia); RPF (Cyprus); SENESCYT (Ecuador); MoER, ERC IUT, and ERDF (Estonia); Academy of Finland, MEC, and HIP (Finland); CEA and CNRS/IN2P3 (France); BMBF, DFG, and HGF (Germany); GSRT (Greece); OTKA and NIH (Hungary); DAE and DST (India); IPM (Iran); SFI (Ireland); INFN (Italy); MSIP and NRF (Republic of Korea); LAS (Lithuania); MOE and UM (Malaysia); BUAP, CINVESTAV, CONACYT, LNS, SEP, and UASLP-FAI (Mexico); MBIE (New Zealand); PAEC (Pakistan); MSHE and NSC (Poland); FCT (Portugal); JINR (Dubna); MON, RosAtom, RAS, RFBR and RAEP (Russia); MESTD (Serbia); SEIDI, CPAN, PCTI and FEDER (Spain); Swiss Funding Agencies (Switzerland); MST (Taipei); ThEPCenter, IPST, STAR, and NSTDA (Thailand); TUBITAK and TAEK (Turkey); NASU and SFFR (Ukraine); STFC (United Kingdom); DOE and NSF (USA).

\hyphenation{Rachada-pisek} Individuals have received support from the Marie-Curie program and the European Research Council and Horizon 2020 Grant, contract No. 675440 (European Union); the Leventis Foundation; the A. P. Sloan Foundation; the Alexander von Humboldt Foundation; the Belgian Federal Science Policy Office; the Fonds pour la Formation \`a la Recherche dans l'Industrie et dans l'Agriculture (FRIA-Belgium); the Agentschap voor Innovatie door Wetenschap en Technologie (IWT-Belgium); the Ministry of Education, Youth and Sports (MEYS) of the Czech Republic; the Council of Science and Industrial Research, India; the HOMING PLUS program of the Foundation for Polish Science, cofinanced from European Union, Regional Development Fund, the Mobility Plus program of the Ministry of Science and Higher Education, the National Science Center (Poland), contracts Harmonia 2014/14/M/ST2/00428, Opus 2014/13/B/ST2/02543, 2014/15/B/ST2/03998, and 2015/19/B/ST2/02861, Sonata-bis 2012/07/E/ST2/01406; the National Priorities Research Program by Qatar National Research Fund; the Programa Clar\'in-COFUND del Principado de Asturias; the Thalis and Aristeia programs cofinanced by EU-ESF and the Greek NSRF; the Rachadapisek Sompot Fund for Postdoctoral Fellowship, Chulalongkorn University and the Chulalongkorn Academic into Its 2nd Century Project Advancement Project (Thailand); and the Welch Foundation, contract C-1845.

\end{acknowledgments}
\clearpage
\bibliography{auto_generated}

\cleardoublepage \appendix\section{The CMS Collaboration \label{app:collab}}\begin{sloppypar}\hyphenpenalty=5000\widowpenalty=500\clubpenalty=5000\textbf{Yerevan Physics Institute,  Yerevan,  Armenia}\\*[0pt]
A.M.~Sirunyan, A.~Tumasyan
\vskip\cmsinstskip
\textbf{Institut f\"{u}r Hochenergiephysik,  Wien,  Austria}\\*[0pt]
W.~Adam, F.~Ambrogi, E.~Asilar, T.~Bergauer, J.~Brandstetter, E.~Brondolin, M.~Dragicevic, J.~Er\"{o}, M.~Flechl, M.~Friedl, R.~Fr\"{u}hwirth\cmsAuthorMark{1}, V.M.~Ghete, J.~Grossmann, J.~Hrubec, M.~Jeitler\cmsAuthorMark{1}, A.~K\"{o}nig, N.~Krammer, I.~Kr\"{a}tschmer, D.~Liko, T.~Madlener, I.~Mikulec, E.~Pree, D.~Rabady, N.~Rad, H.~Rohringer, J.~Schieck\cmsAuthorMark{1}, R.~Sch\"{o}fbeck, M.~Spanring, D.~Spitzbart, J.~Strauss, W.~Waltenberger, J.~Wittmann, C.-E.~Wulz\cmsAuthorMark{1}, M.~Zarucki
% ...
\vskip\cmsinstskip
% ...
1:~~Also at Vienna University of Technology, Vienna, Austria\\
% ...

\end{sloppypar}
\end{document}